# DigiCam - Fully Digital Compact Read-out and Trigger Electronics for the SST-1M Telescope proposed for the Cherenkov Telescope Array

**P. Rajda**[k][1], **K. Ziętara**[b], **W. Bilnik**[k], **J. Błocki**[g], **L. Bogacz**[e], **T. Bulik**[d], **F. Cadoux**[a],
**A. Christov**[a], **M. Curyło**[g], **D. della Volpe**[a], **M. Dyrda**[g], **Y. Favre**[a], **A. Frankowski**[c],
**Ł. Grudnik**[g], **M. Grudzińska**[d], **M. Heller**[a], **B. Idźkowski**[b], **M. Jamrozy**[b], **M. Janiak**[c],
**J. Kasperek**[k], **K. Lalik**[k], **E. Lyard**[f], **E. Mach**[g], **D. Mandat**[m], **A. Marszałek**[h,b],
**J. Michałowski**[g], **R. Moderski**[c], **M. Rameez**[a], **T. Montaruli**[a], **A. Neronov**[f],
**J. Niemiec**[g], **M. Ostrowski**[b], **P. Paśko**[h], **M. Pech**[m], **A. Porcelli**[a], **E. Prandini**[f],
**E. jr Schioppa**[a], **P. Schovanek**[m], **K. Seweryn**[h], **K. Skowron**[g], **V. Sliusar**[i],
**M. Sowiński**[g], **Ł. Stawarz**[b], **M. Stodulska**[b], **M. Stodulski**[g], **S. Toscano**[f,l],
**I. Troyano Pujadas**[a], **R. Walter**[f], **M. Więcek**[k], **A. Zagdański**[b], **P. Żychowski**[g]

[a] *DPNC – Université de Genève, Genève, Switzerland;* [b] *Astronomical Observatory, Jagiellonian University, Kraków, Poland;* [c] *Nicolaus Copernicus Astronomical Centre, Polish Academy of Sciences, Warsaw, Poland;* [d] *Astronomical Observatory, University of Warsaw, Warsaw, Poland;* [e] *Department of Information Technologies, Jagiellonian University, Kraków, Poland;* [f] *ISDC, Observatoire de Genève, Université de Genève, Versoix, Switzerland;* [g] *Instytut Fizyki Jądrowej im. H. Niewodniczańskiego Polskiej Akademii Nauk, Kraków, Poland;* [h] *Centrum Badań Kosmicznych Polskiej Akademii Nauk, Warsaw, Poland;* [i] *Astronomical Observatory, Taras Shevchenko National University of Kyiv, Kyiv, Ukraine;* [k] *AGH University of Science and Technology, Kraków, Poland;* [l] *Vrije Universiteit Brussels, Brussels, Belgium;* [m] *Institute of Physics of the Czech Academy of Sciences, Prague, Czech Republic*

**for the CTA Consortium** (full consortium author list at http://cta-observatory.org)

*e-mail:* `pjrajda@agh.edu.pl`

The SST-1M is one of three prototype small-sized telescope designs proposed for the Cherenkov Telescope Array, and is built by a consortium of Polish and Swiss institutions. The SST-1M will operate with DigiCam - an innovative, compact camera with fully digital read-out and trigger electronics. A high level of integration will be achieved by massively deploying state-of-the-art multi-gigabit transmission channels, beginning from the ADC flash converters, through the internal data and trigger signals transmission over backplanes and cables, to the camera's server link. Such an approach makes it possible to design the camera to fit the size and weight requirements of the SST-1M exactly, and provide low power consumption, high reliability and long lifetime. The structure of the digital electronics will be presented, along with main physical building blocks and the internal architecture of FPGA functional subsystems.

---

[1]Speaker

## 1.    Introduction

The single-mirror Small Size Telescope (SST-1M) prototype is currently being developed by a consortium of Polish and Swiss institutions. The construction is based on the proven Davies-Cotton (DC) design, used in some of the currently operated VHE gamma-ray observatories [1]. The new idea is to equip the telescope with a fully digital camera, based on silicon photomultipliers (SiPM).

## 2.    Camera architecture

The telescope camera is based on the FlashCam concept [2] and bears the name DigiCam [3]. The design separates the photon detection plane (PDP) from the digital camera electronics (Figure 1). With such an approach, the mechanical construction of the PDP [4] follows the needs of physics and optics (hexagonal geometry) and the construction of the digital readout and trigger systems follows the needs for a compact, standard and easily cooled electronics. However, both parts will be fit into one, compact enclosure, mounted on top of the telescope mast.

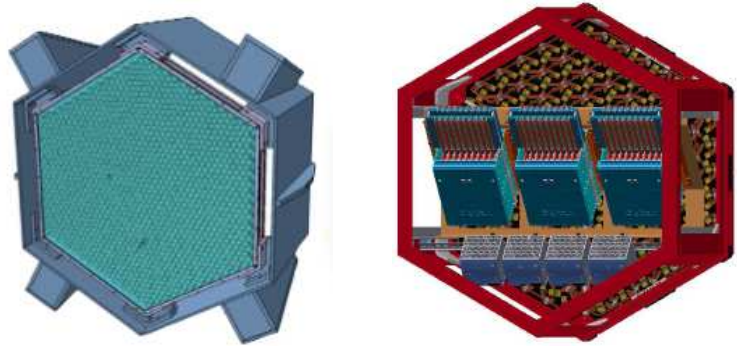

Figure 1. The hexagonal PDP of the camera (left) and the frame for the crates (right).

### 2.1    Photo Detection Plane

In the camera, the signal coming from the photodetector, after amplification and shaping, is digitized and both the trigger decisions and the readout is done on digital signals. Such a scheme allows for a great flexibility in trigger algorithms and readout organization. The choice of using SiPMs instead of vacuum tube photomultipliers (PMTs) has been proved to be worth by their successful use in the camera of the FACT telescope taking data in La Palma [5]. Due to much smaller dimensions, SiPMs allowed for shrinking the size of the whole telescope, but prevented the use of original FlashCam design, which was then thoroughly redesigned to increase the integration level of the digital part and its compactness. The SST-1M camera is now a hexagon of about 1 m flat-to-flat wide and about 60 cm thick, with a PDP flat-to-flat diameter of 88 cm.

### 2.2    Digital electronics - hardware

The digital subsystem hardware will consist of three so-called "microcrates" (Figure 2 – right), each containing the same set of boards and serving for the one sector of PDP (1/3 of area = 432 pixels, Figure 2 – left). A single microcrate will include 9 ADC boards and 1 trigger board. Each ADC board will process 48 channels (pixels), coming from 4 PDP modules. The trigger board will process trigger signals from all 9 ADC boards within the microcrate and some extra trigger data from the neighboring pixels from remaining two sectors (other microcrates). Thus, 27 ADC and 3 trigger boards will be required to handle signals from all 1296 pixels.

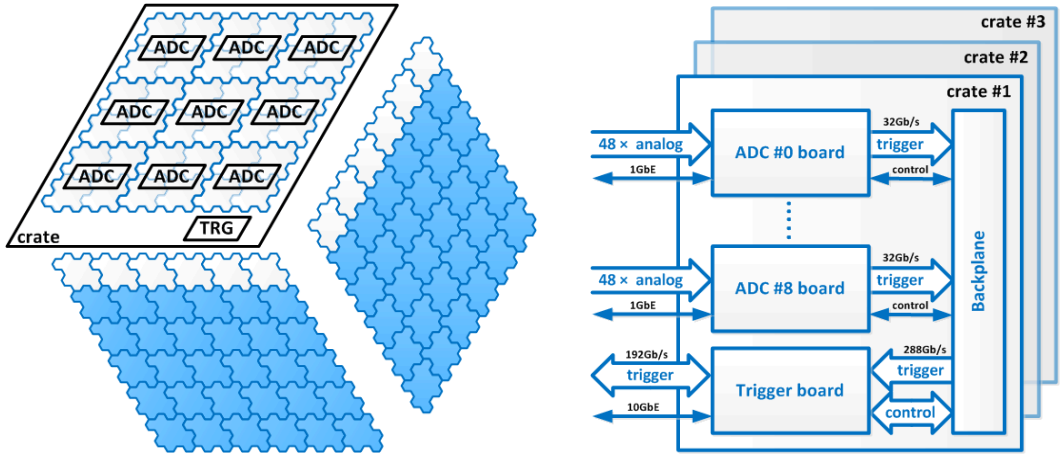

Figure 2. Organization of the photon detection plane (left) and architecture of the digital electronics subsystem (right).

### 2.2.1 ADC board

The ADC board main tasks are: to digitize the analog signals of the PDP and to pre-process digitized signals and store them in digital ring buffers. Its job is also to calculate the first level trigger (L0) signals and send them continuously to the trigger board. Since the main goal of this new design is to increase the level of integration, using the single-channel ADCs with parallel data output [2] was not possible. It would take too large area of PCBs and require a high number of routing layers and FPGA pins. Thus another concept was chosen – to employ highly integrated, multi-channel A/D converters, equipped with the fast serial digital data interfaces. Such a data interface for a 12-bit converter needs only 1 pair of signals, instead of 6 pairs required by parallel output. Using such devices releases constraints on number of PCB layers and FPGA pins. The authors decided to test two types of flash converters (FADC): 4-channel AD9239 converter from Analog Devices and 2-channel ISLA222S25 converter from Intersil, both working at 12-bit, 250MS/s.

The block diagram of the ADC board is shown in Figure 3, and more datailed description may be found in [3]. The layout of ADC board based on ISLA222S25 is shown in Figure 5, left.

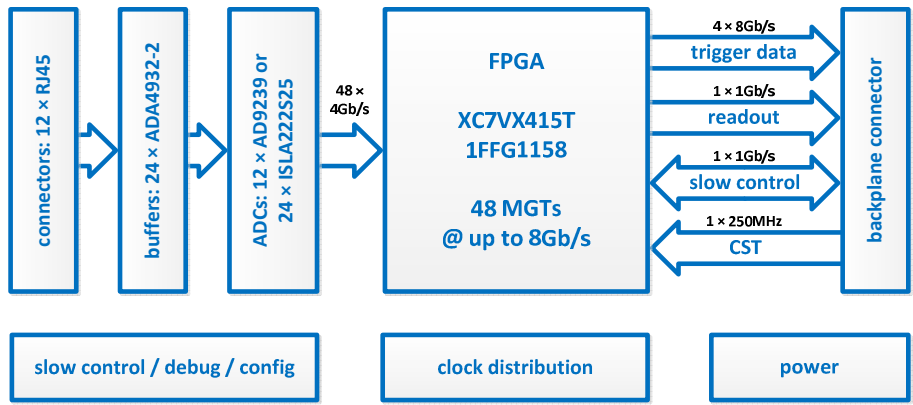

Figure 3. Block diagram of the ADC board.

*DigiCam - Fully Digital Compact Electronics for the SST-1M CTA Telescope*

## 2.2.2 Trigger board

The trigger board main tasks are: to receive the trigger signals from all the ADC boards within the crate (sector) and from the neighboring channels in other sectors, to calculate the second level trigger (L1) signals and send the decision over the whole digital electronics. Its job is also to collect the event data from all the ADC boards within the crate and send them to the central acquisition system. Thus the main functional devices on the board are FPGA, DDR3 memory card and a number of high-speed connectors.

The block diagram of the board is shown in Figure 4, and more datailed description may be found in [3]. The layout of ADC board is shown in Figure 5, right.

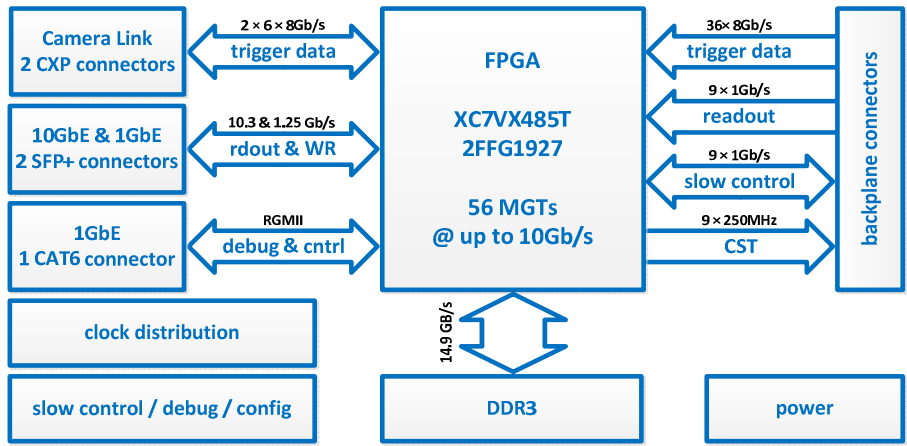

Figure 4. Block diagram of the trigger board.

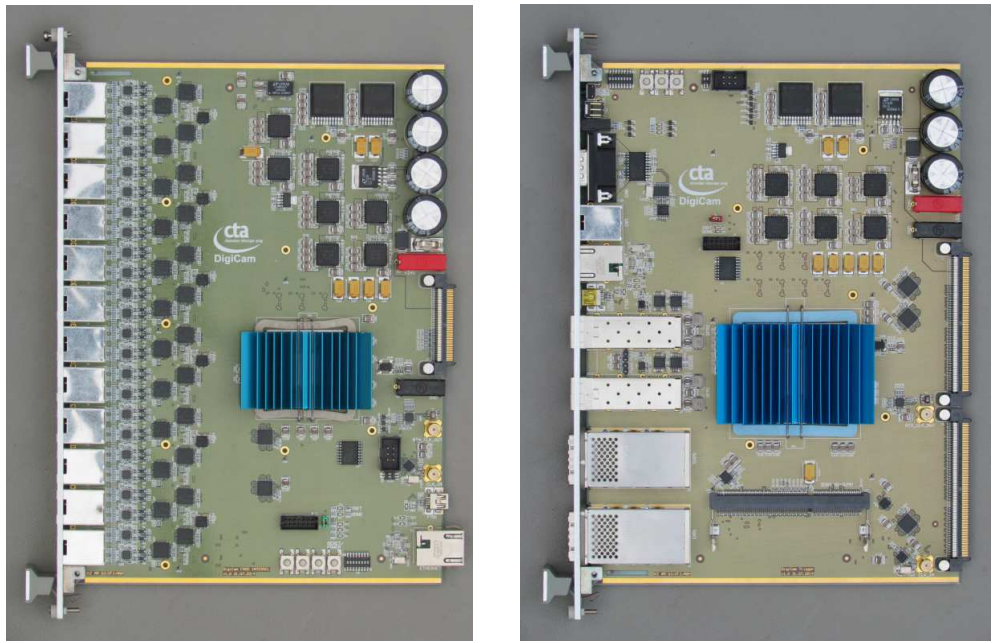

Figure 5. Layout of ADC board (left) and trigger board (right).

### 2.2.3 Camera intraconnect and interconnect

There are two layers of internal connections between modules of the digital part of the camera. First, there is a custom-designed backplane within each crate, providing the connectivity between the trigger and the ADC boards. Second, there are high-throughput cables connecting all three crates, each with all the others, to interchange data for the whole camera.

The backplane is a multi-layer PCB with connectors for 10 boards (9 ADC boards and 1 trigger board), and connections between them organized with a star topology, with the trigger board being the logical center (physically the trigger board is placed at backplane's end). To interconnect the camera's crates there will be high-throughput, multi-channel Camera Link used, based on InfiniBand CXP 84-pin cables and connectors, each including 24 pairs ready for 12.5Gbps transmission. Every trigger board contains two CXP connectors, allowing to form a triangle interconnect topology.

There are different kinds of interfaces and protocols needed for the interconnecting the camera with central acquisition and control system, placed outside of the telescope. The highest throughput is required for event data readout. One up to three 10Gb Ethernet links (one per camera or one per each crate) are provided. For the slow control of camera resources (PDP, power supply, conditioning, etc.) a 1Gb Ethernet will be sufficient. There will be also a need for connectivity resources required to synchronize the telescope's camera with cameras in other telescopes. For all these purposes authors decided to use only the optical fibers media.

More datailed description of connection resources may be found in [3].

### 2.3 Digital electronics – firmware

For designed operation of the programmable FPGA-based boards, a number of digital subsystems have to be developed, along with control software running on the internal MicroBlaze embedded soft-processor. All the below-described subsystems were designed using a Xilinx Vivado software suite. Additionally, some extra software for PC had to be developed to provide a high-level control, as well as big data sets transmission (FPGA reconfiguration) and reception (captured data read-out).

### 2.3.1 MicroBlaze framework

The digital system design of each card is based on MicroBlaze soft processor framework composed of set of standard and custom made IP-cores interconnected together using AXI4 and AXI4 Stream buses. Processor is responsible for initialization, control and monitoring of all the hardware components and interfaces used. It provides debugging and internal state monitoring using JTAG interface and MDM module, as well as dedicated USB serial port and 1Gbit Ethernet interface, which can be used alternatively as a control subsystem port. The firmware written in C is based on standalone, single-threaded environment without operating system.

### 2.3.2 FADC readout sybsystem

The readout of the FADC converters is implemented using only receiver sections of high speed serial transceivers (MGT) available in Virtex 7 FPGA. Converters transmit the acquired samples in serial synchronous frames with the speed of either 3.75Gbps or 4Gbps, depending on the converter type finally chosen. There are also two different frame formats used for this purpose, based on either JESD204 standard with 8B/10B coding or custom protocol with scrambling. At the initial state the readout subsystem performs synchronization procedure to find the alignment of the incoming frames. Then consecutive frames are received, decoded and extracted samples are stored in the ring buffers of length up to 1024 samples (4us), where they wait for the trigger algorithm decision. In case of positive trigger decision, subsystem reads out block of consecutive samples from the ring buffer and sends it to the readout buffer. Size of the block transferred as well as its relative time position to the trigger signal is programmable. For testing and quality monitoring of the readout channels the subsystem implements Bit Error Rate measurement functionality utilizing Pseudo Random Binary Sequence patterns transmitted by the converters.

### 2.3.3 Trigger calculation subsystem

The trigger algorithm is a highly parallelized, systolic computing structure, that calculates the event occurrence based on exceeding the threshold on the sum of L0 trigger data, counted for 7- or 19-triplet patches (neighborhoods). Figure 6 shows the single trigger board calculation area. Each hexagon on this picture stands for one trigger triplet (minimum portion of trigger data), derived from 3 detectors in ADC board. The central area is occupied by data from 9 ADC boards and provide 144 triplets (local). Bottom and right borders are formed by the overlapping data coming from neighbouring crates, and the rest of the area is zero-filled (phantom) to enable easy algorithmic trigger description / calculation. Thus, the whole area is formed in total of 256 triplets.

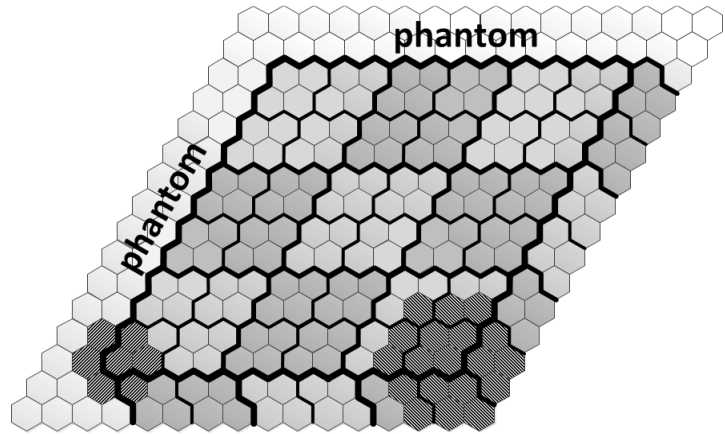

Figure 6. Single trigger board calculation area with 7-triplet (left bottom hexagon) and 19-triplet (right bottom hexagon) patches marked.

Trigger calculation is done only for patches having the center within a camera sector (144). There are 7-triplet and 19-triplet patches marked for example in Figure 6. The 7-triplets neighborhood trigger algorithm implemented on this platform utilized only 2% of the FPGA resources, being able to be run twice faster than required. Such a surplus well proves the intended ability of the proposed architecture to implement other real-time trigger and/or data processing algorithms.

### 2.3.4 Trigger data delay equalization subsystem

Due to the different connection architecture, all the received trigger signals need to be synchronized to present the data from the same ADC sampling point in time. For this purpose a special channel bonding procedure must be performed during the camera startup to equalize delays between transmission channels. After the bonding, the trigger data from consecutive sampling points are presented to the trigger algorithm properly. The main reason for a trigger signals delay skew are differences in the paths of the trigger data between local and neighbour fields. Local field signals exploit only one MGT transmission (ADC Board ⇨ Trigger Board), and the neighbour / overlap area data need two consecutive MGT transmissions to be finally presented at destination (ADC Board ⇨ Trigger Board ⇨ neighbour Trigger Board). Since a delay of each transfer is about 100ns (~96 ns per transceiver with elastic buffer and only 4 ns per cable), it was assumed that all the data will be effectively delayed by a fixed amount of time, e.g. 256 ns (64 sample periods). For this purpose a delay skew compensation module has been designed, that is closely coupled with original MGT quad module and allows for timing compensation of data signal transmission.

### 2.3.5 10GbE data read-out link

The 10Gb Ethernet link was tested with an FPGA design, that funnels nine read-out data transmission sources, each running at 1Gbps, into one channel feeding the 10GbE interface IPcore. All the sources provided the data on AXI stream interfaces to the AXI interconnect, working in a round-robin arbitration mode. Transmitted data were received to the RAM memory of the server machine, equipped with 10GbE interface. The data rate achieved 9Mbps, which is well over the needs of single camera crate. Thus, the number of 10GbE interfaces within a camera server required to connect a read-out electronics may be reduced from three to one, saving some costs. In such a case only one of the microcrates has to be equipped with optical transceiver, and it should gather the read-out data from remaining two crates via readout data interchange channel within the Camera Link.

### 2.3.6 FPGA reconfiguration subsystem

To increase a reliability, a life-time and a manageability of the whole design, FPGA reconfiguration management solutions were proposed. The reconfiguration resources are based on external Flash memories, that store the configuration data streams on each board containing the FPGA. While using the 512Mb memory devices, each may store up to three different configurations. One of them is set as a 'golden revision' and stored in a protected area of the memory, with no permission for further writing. Remaining area may store two 'multi-boot revisions', which may be easily changed from outside of the camera. Single uncompressed FPGA configuration set for ADC board contains almost 138 megabits of data, and the dataset for Trigger board – over 162 megabits. Writing such a big data set to the Flash memory via an 1GbE maintenance link lasts only for 90 seconds with 62.5MHz Flash memory clock and 4-bit SPI data bus. To improve the time required to upgrade the whole camera, the broadcast mode will be employed, allowing to provide the configuration data to all the boards of a kind at once.

## 3. Conclusion

No specific limitation to meet the requirements of the CTA project has been found for a 4m Davies-Cotton telescope equipped with a fully digital camera. Such an electronics has many advantages: it is lightweight, easy to transport and install, easy to operate, has a high reliability and time-of-life. Now the boards are fully tested (Figure 7) and are being replicated to form a whole camera electronics. Concurrently, the ongoing firmware development is focused on integration of all functional subsystems. After this the whole camera (PDP and read-out electronics) will be integrated and functionally tested in the lab. Then it will be mounted on the telescope structure and subjected to environmental tests.

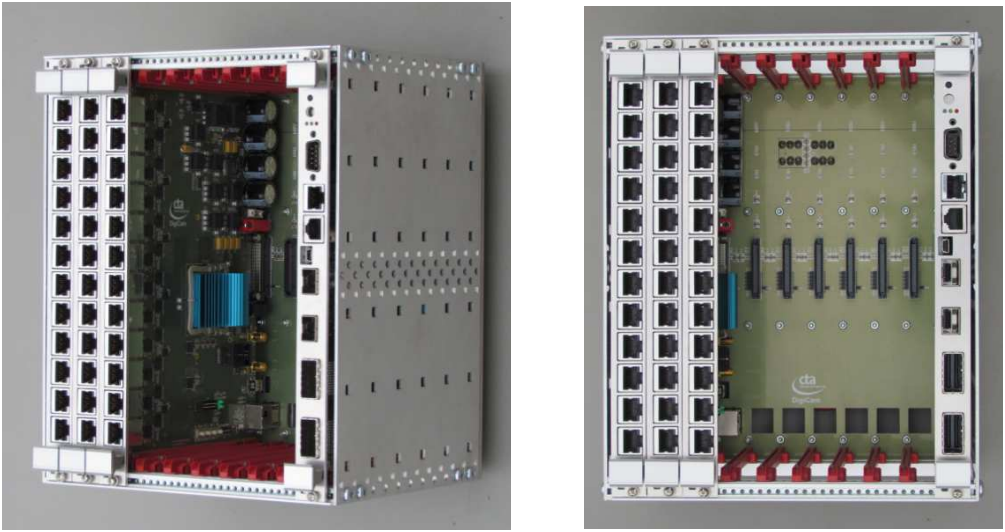

Figure 7. Partially armed microcrate, showing the backplane (right)

## Acknowledgments

We gratefully acknowledge support from the agencies and organizations listed under Funding Agencies at this website: http://www.cta-observatory.org/. In particular we are grateful for support from the Polish NCN grant DEC-2011/01/M/ST9/01891, the Polish MNiSW grant no. 498/1/FNiTP/FNiTP/2010, the University of Geneva, and the Swiss National Foundation.